\documentclass[twocolumn]{svjour3}

\usepackage[pdftex]{graphicx}
\usepackage{amsmath}
\usepackage[colorlinks,
            unicode,
            linkcolor=black,
            filecolor=blue,
            citecolor=blue,
            urlcolor=blue]{hyperref}
\usepackage[center]{caption}

\begin{document}


\title{Inertia compensation while scanning screw threads on coordinate-measuring machines}

\author{Sergey~Kosarevsky* \and Viktor Latypov}

\institute{* Corresponding author.\\
           S.V. Kosarevsky \at
           Faculty of Technology, \\
           Saint-Petersburg Institute of Mechanical-Engineering, \\ Saint-Petersburg, Russia 195197 \\
           \email{kosarevsky@linderdaum.com}  
           \and
           V.N. Latypov \at
           Faculty of Applied Mathematics and Control Processes, \\
           Saint-Petersburg State University, \\ Saint-Petersburg, Russia 198504
}

\date{Received: date / Accepted: date}

\maketitle               	

\begin{abstract}
Usage of scanning coordinate-measuring machines for inspection of screw threads has become a common practice nowadays.
Compared to touch trigger probing, scanning capabilities allow to speed up measuring process while still maintaining high accuracy.
However, in some cases accuracy drasticaly depends on the scanning speed. In this paper a compensation method is proposed allowing to 
reduce the influence of some dynamic effects while scanning screw threads on coordinate-measuring machines.
\end{abstract}


\maketitle

\section{Introduction}

\indent
Numerical compensation of measurement errors is a well developed field of dimensional metrology \cite{GUM}.
This type of compensation originates from different CAA (computer-aided accuracy) methods and is
extensively used in modern CMMs and CNC machines during recent years \cite{Trapet97}, \cite{Jenq95}.

\indent
To successfully compensate measurement results one should clearly undertand the exact source of errors.
Measuring uncertainty on coordinate-measuring machines depends on many factors. They can be summarized
into the following types \cite{Schwenke08}:

\begin{itemize}
   \item kinematic;
   \item thermo-mechanical;
   \item loads;
   \item dynamic;
   \item control software.
\end{itemize}

\indent
Some dynamic effects produce non-constant measurement errors varying according to known models.
These models depend entirely upon specific measurement task and development of generic model is
troublesome. One of the widely accepted example of such kind of model is pretravel 
compensation \cite{Estler96}, \cite{Woz03-1}, \cite{Woz03-2}.

\indent
In this paper a novel compensation method is given to reduce minor and major diameters measurement error from
dynamic effects emerging while scanning screw threads.

\section{Related work}

\indent
Due to inertial forces while high-speed scanning freeform workpieces with low radius of curvature on coordinate-measuring 
machines it is difficult for the control software to maintain stable contact between 
the probe and the surface. This is one of the reasons preventing accurate scanning at high speeds \cite{Pereira07}.
The study of dynamic effects was conducted by many researchers. 
Application of signal analysis and processing theory to dimensional metrology was studied in \cite{Hessling08}. 
Pereira and Hocken \cite{Pereira07} proposed classification and compensation methods for dynamic
errors of scanning coordinate-measuring machines. They used Taylor series and Fourier analysis to compensate measurement of 
circular features.
ISO 10360 \cite{ISO10360} defines acceptance tests for scanning coordinate-measuring machines.
Farooqui and Morse \cite{Far07} proposed reference artifacts and tests to compare scanning performance of
different coordinate-measuring machines.
Szelewski et al. \cite{Szel07} conducted experimental research and concluded that freeform surfaces scanning 
time reduction is limited at high speeds by acceleration and deceleration of the probing system.

\section{Theoretical model of probe inertial motion}

\begin{figure}[!ht]
   \begin{center}
      \includegraphics[width=6cm]{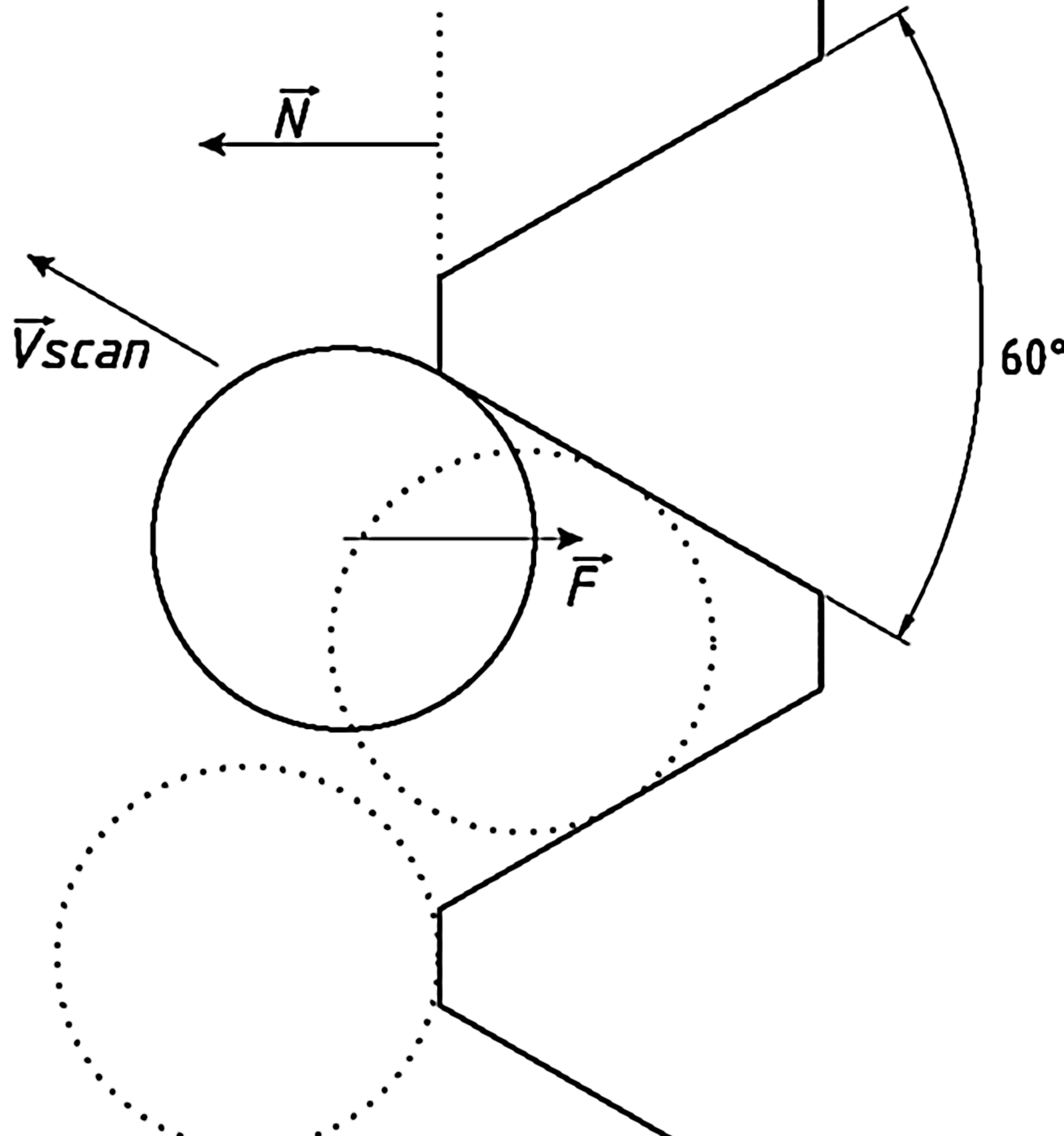}
   \end{center}
   \caption{Scanning thread surface with a spherical probe.}
   \label{Sketch}
\end{figure}

\indent
High-speed longitudinal scanning of a thread surface results in loss of contact between the probe and the thread surface.
Because of this coordinate-measuring machine will measure points not belonging to the thread surface. The actual value of the
minor diameter (for internal threads) or the major diameter (for external threads) can be distorted. This effect is 
noticeable at scanning speeds above 5~mm/s and the probing system mass of over 0.1~kg.

\indent
The ``helix'' scanning strategy is free from this effect since no surface part with low 
radius of curvature is scanned at high speed.

\begin{figure}[!ht]
   \begin{center}
      \includegraphics[width=8.5cm]{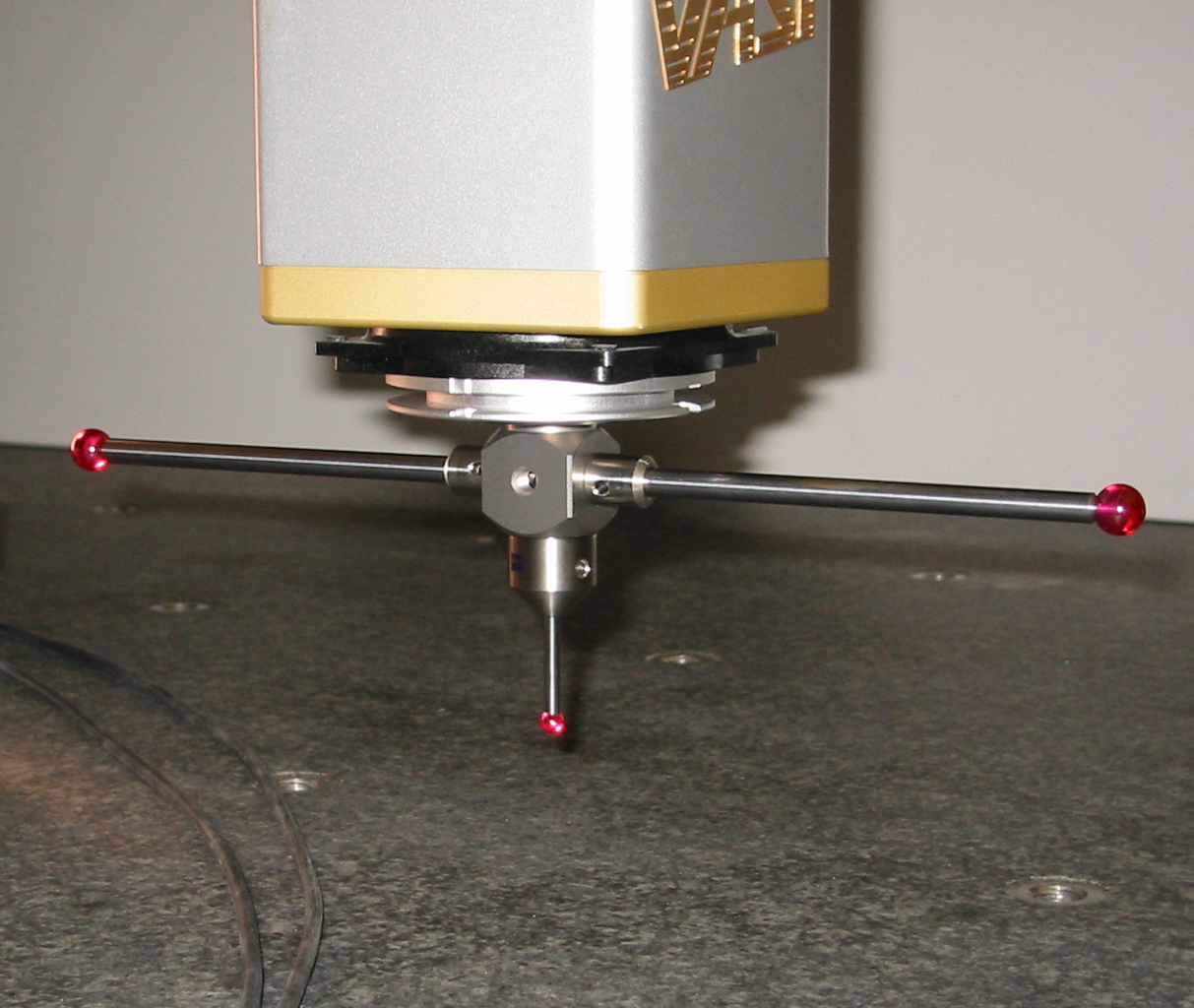}
   \end{center}
   \caption{Probing system floating inside the CMM probing head.}
   \label{ProbingSystem}
\end{figure}

\indent
Energy conservation yields (fig.~\ref{Sketch}):
\begin{equation}
   \frac{m \cdot \left( \vec{V_{scan}} \cdot \vec{N} \right)^{2} }{2} = | \vec{F} | \cdot \delta,
   \label{EnergyConservation}
\end{equation}

\noindent
where $ m $ is the mass of the probing system floating inside the CMM probing head (fig.~\ref{ProbingSystem}), kg; 
$ \vec{V_{scan}} $ is the scanning speed, m/s;
$ \vec{F} $ is the contact force, N; 
$ \vec{N} $ is the unit vector othogonal to the thread axis;
$ \delta $ is the separation of the probe from the surface, m.

\indent
Assuming that \vec{N} and \vec{F} are collinear the separation value $ \delta $ can be derived 
from the equation~(\ref{EnergyConservation}):
\begin{equation*}
   \delta = \frac{ m \cdot V_{scan}^{2} \cdot \cos^{2}\frac{\alpha}{2} }{ 2 \cdot F },
\end{equation*}

\noindent
where $ \alpha $ is the thread angle.

\indent
Since the separation $ \delta $ is axisymmetric (fig.~\ref{Diameters}) the diameter measurement error is
\begin{equation}
   D_{\delta} = \frac{ m \cdot V_{scan}^{2} \cdot \cos^{2}\frac{\alpha}{2} }{ F }.
   \label{EqDdelta}
\end{equation}

\section{Compensation}

\begin{figure}[!ht]
   \begin{center}
      \includegraphics[width=6cm]{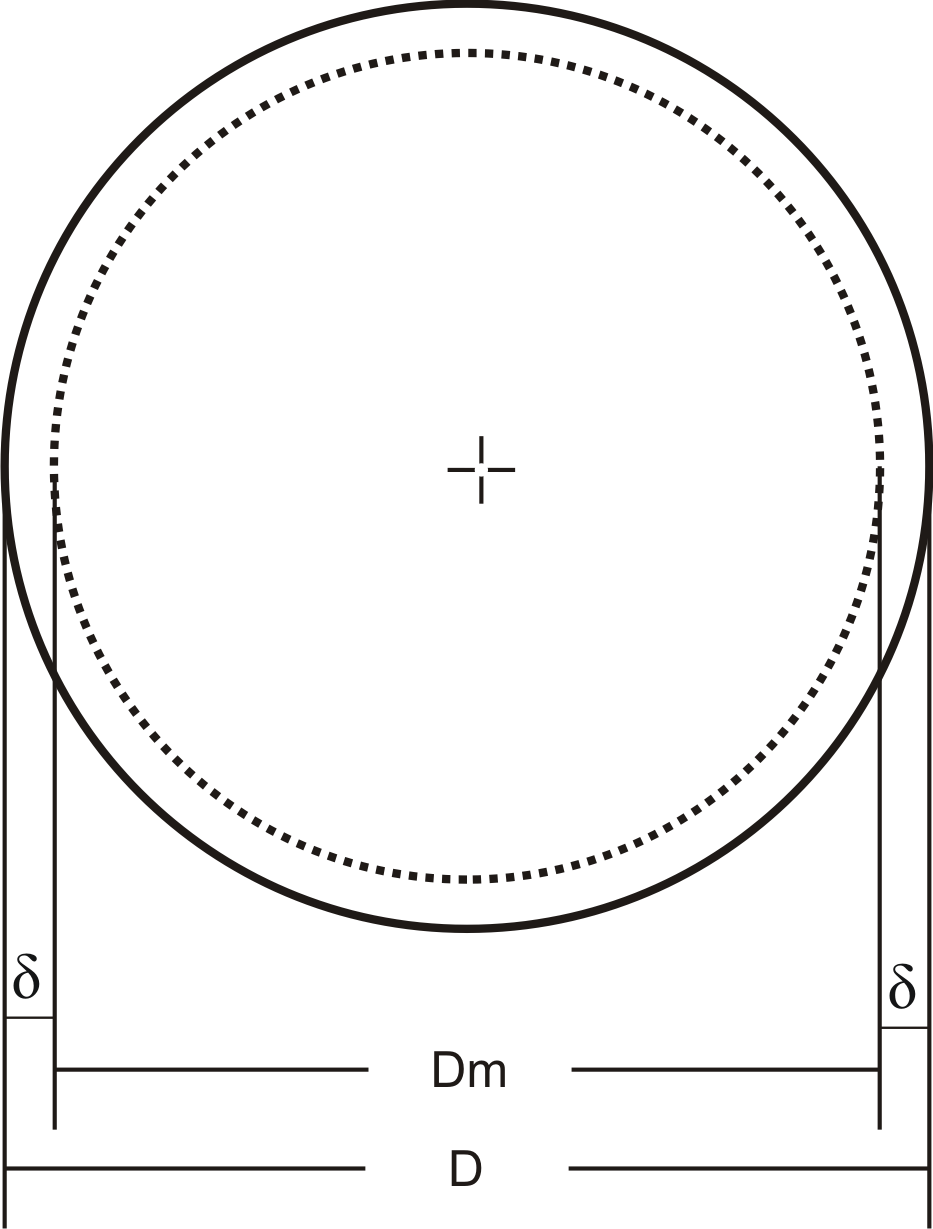}
   \end{center}
   \caption{Compensated and measured diameters.}
   \label{Diameters}
\end{figure}

\indent
The equation (\ref{EqDdelta}) can be used to obtain the diameter compensation formula:
\begin{equation*}
   D = D_{m} \pm \frac{m \cdot V^{2}_{scan} \cdot \cos^{2}{\frac{\alpha}{2}} }{F},
\end{equation*}

\noindent
where $ D $ is the compensated diameter, m (fig.~\ref{Diameters}); 
$ D_{m} $ is the measured diameter, m. The upper sign applies to internal threads and the lower sign to 
external threads respectively.

%
%
%
%
%
                                                                                    
\section{Experimental verification of the model}

\indent
We scanned the ISO metric screw thread ring gauge M170x6 \cite{ISO68} to measure its minor diameter. The diameter
was calculated as the maximal inscribed cylinder with both filtering and outlier elimination disabled.
The mass of our probing system was 0.1~kg according to manufacturer's specification, contact force was set in CMM 
software to 0.2~N.

\indent
Theoretical and experimental values for different scanning speeds are summarized in table~\ref{ResultsTable}.

\indent
Experimental values were obtained on a Carl Zeiss PRISMO 10 S-ACC ($ MPE_{E} = 1.7 + \frac{L}{350} $) coordinate-measuring machine.
The value of the minor diameter scanned at 1 mm/s was assumed to be the actual value ($ D_{\delta.Exp.} = 0 $).
Each of the resulted experimental errors were averaged by 10 measurements.

\begin{table*}
\caption{Effect of scanning speed on measurement results.}
\label{ResultsTable}
\begin{center}
\begin{tabular}{cccccc}
\hline
Speed, mm/s  &  Separation $ \delta $, $ \mu m $  &  $ D_{\delta} $, $ \mu m $ &  $ D_{\delta.Exp.} $, $ \mu m $  &  $ D_{\delta.Exp.} - D_{\delta} $, $ \mu m $  & Reduction, \%  \\
\hline
1  &  0.2 	&   0.4 	&   0.0 	&   -0.4 	&   0.0   \\
2  &  0.8 	&   1.5 	&   0.4 	&   -1.1 	&   26.7  \\
3  &  1.7 	&   3.4 	&   1.8 	&   -1.6 	&   53.0  \\
4  &  3.0 	&   6.0 	&   2.0 	&   -4.0 	&   33.3  \\
5  &  4.7 	&   9.4 	&   3.1 	&   -6.3 	&   33.0  \\
6  &  6.8 	&   13.5 	&   11.0 	&   -2.5 	&   81.5  \\
7  &  9.2 	&   18.4 	&   11.9 	&   -6.5 	&   64.7  \\
8  &  12.0 	&   24.0 	&   31.6 	&   7.6 	&   68.3  \\
9  &  15.2 	&   30.4 	&   37.6 	&   7.2 	&   76.3  \\
10 &  18.8 	&   37.5 	&   46.1 	&   8.6 	&   77.1  \\
11 &  22.7 	&   45.4 	&   38.5 	&   -6.9 	&   84.8  \\
12 &  27.0 	&   54.0 	&   40.6 	&  -13.4 	&   75.2  \\
13 &  31.7 	&   63.4 	&   80.2 	&   16.8 	&   73.5  \\
14 &  36.8 	&   73.5 	&   90.8 	&   17.3 	&   76.5  \\
15 &  42.2 	&   84.4 	&   123.7 	&   39.3 	&   53.4  \\
\hline
\end{tabular}
\end{center}
\end{table*}

\begin{figure*}[!ht]
   \begin{center}
      \includegraphics[width=15cm]{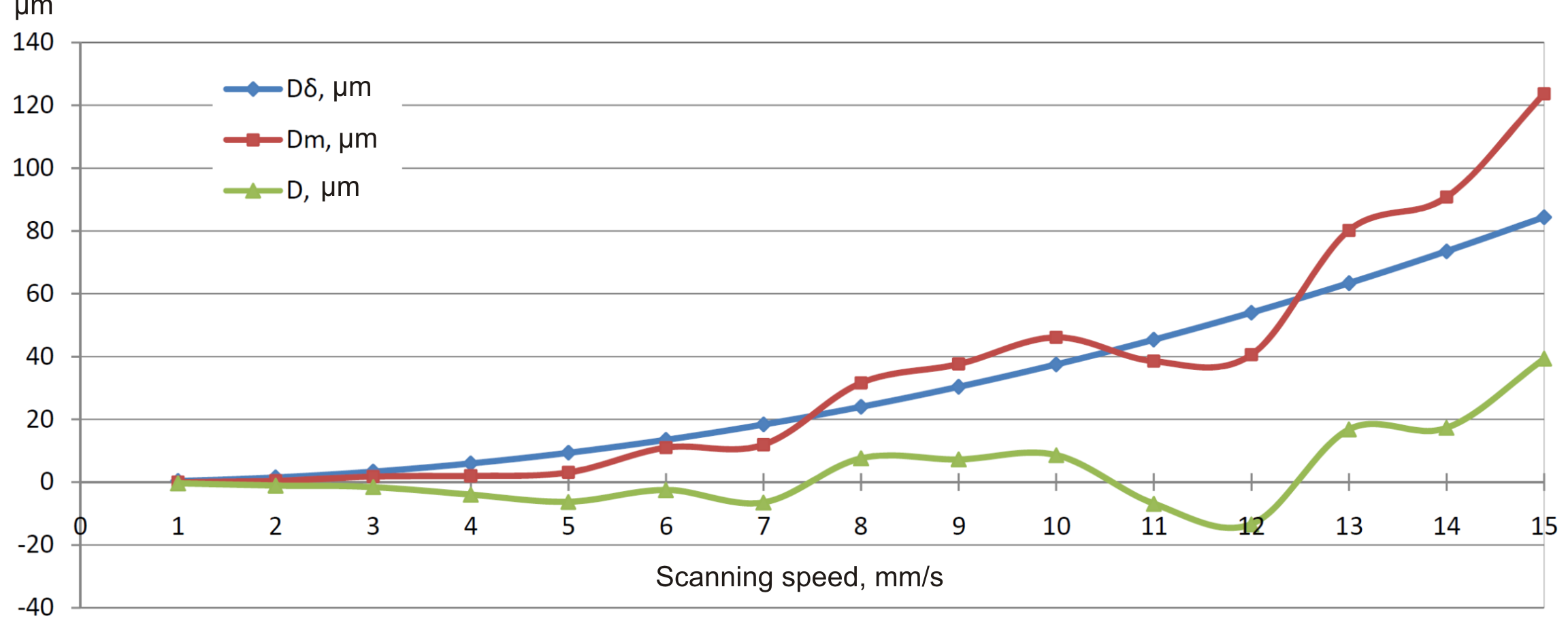}
   \end{center}
   \caption{Measurement error versus scanning speed.}
   \label{ChartMEvsSS}
\end{figure*}

\begin{figure*}[!ht]
   \begin{center}
      \includegraphics[width=15cm]{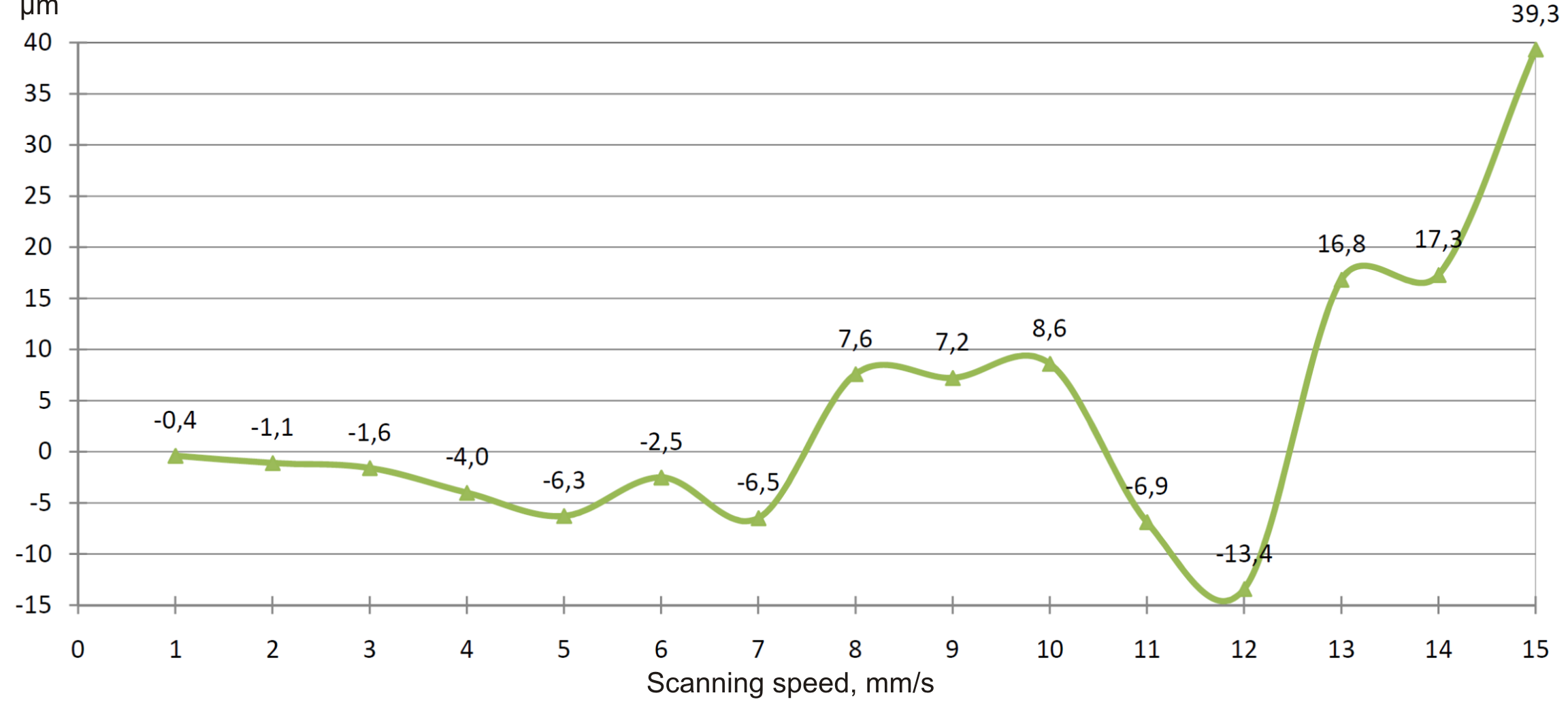}
   \end{center}
   \caption{Compensated measurement error versus scanning speed.}
   \label{ChartCMEvsSS}
\end{figure*}

\indent
Plots of the estimated error ($ D_{\delta} $), the experimental error ($ D_{\delta.Exp.} $) and
the compensated error ($ D_{\delta.Exp.} - D_{\delta} $) are displayed in fig.~\ref{ChartMEvsSS}.
The compensated error is also individually shown in fig.~\ref{ChartCMEvsSS}.

\begin{figure*}[!ht]
   \begin{center}
      \includegraphics[width=15cm]{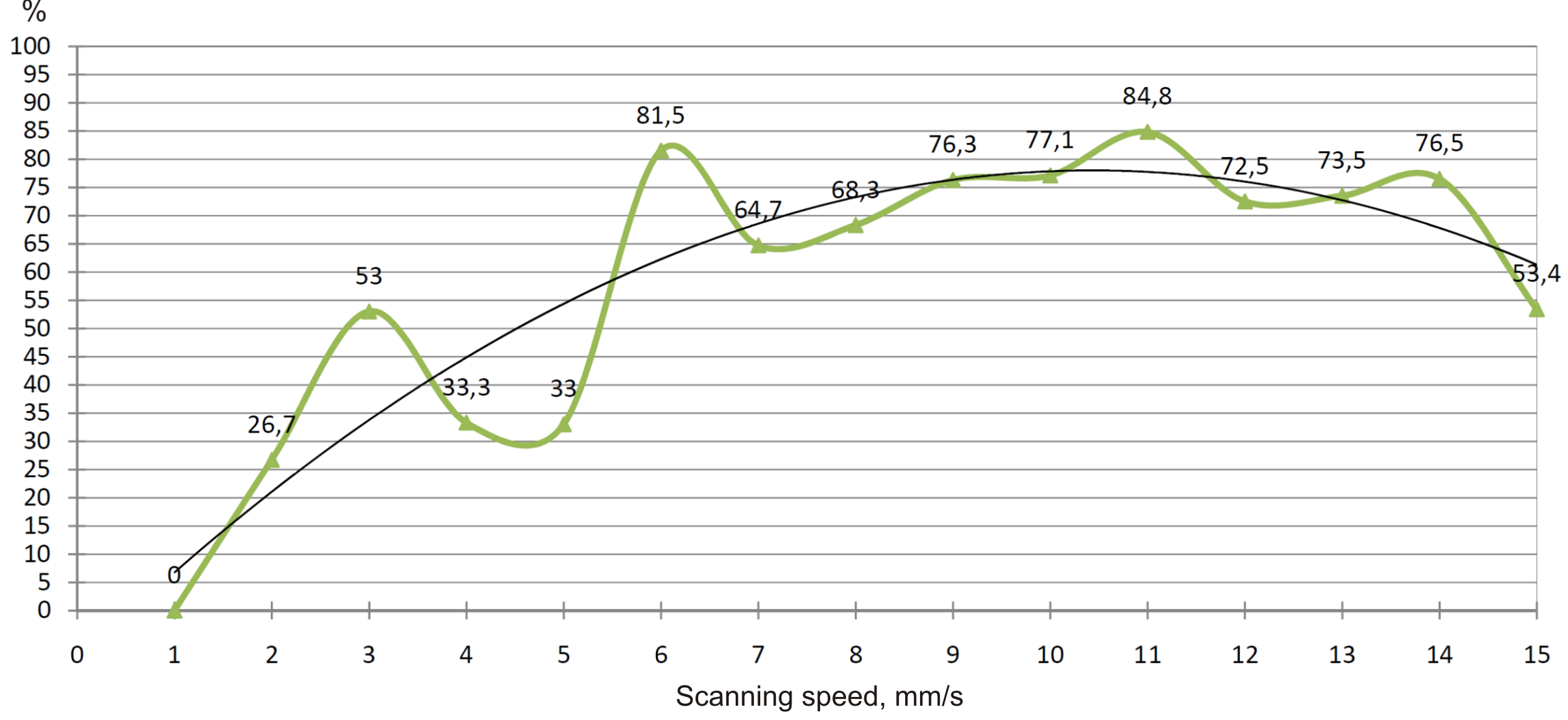}
   \end{center}
   \caption{Error reduction.}
   \label{ChartERvsSS}
\end{figure*}

\indent
The percentage of error reduction was calculated as
\begin{equation*}
   R = \left( 1 - \frac{ | D_{\delta.Exp.} | - D_{\delta} }{ D_{\delta} } \right) \cdot 100\%,
\end{equation*}

\noindent
and is displayed in fig.~\ref{ChartERvsSS}. Quadratic least-squares approximation (black) is shown for reference.

\section{Conclusions}

\indent
The most significant error reduction (fig.~\ref{ChartERvsSS}) can be achieved in the interval of scanning speeds
from 6~mm/s and up to 12~mm/s. At scanning speeds below 5~mm/s the influence of the studied effect is surpassed
by the $ MPE_{E} $ value of the tested CMM and cannot be treated unambiguously (note large fluctuations in
error reduction at speeds below 5~mm/s). At speeds above 12~mm/s the compensated error is quickly 
surpassing 20~$ \mu m $ and can be unaccaptable for the most practical applications.

\indent
Even though the mass of the probing system was taken directly from the manufacturer's specification and no 
actual mass measurement was performed the calculated compensation values has proven to be useful.

\indent
Though numerical compensation of the measurement results is a rapidly growing field of dimensional metrology
its application to dynamic effects compensation is to be extended. Experimental activities described in this paper
confirm the method proposed is able to reduce the influence of interia of the probing system on measurement error
of screw thread's minor and major diameters. This method can be easily implemented directly
in CMM software to allow automatic inertia compensation.

\section{Acknowledgement}

\indent
This work was carried out with the support of the OPTEC --- Carl Zeiss grant 2010, \url{http://optec.zeiss.ru}.










\end{document}